\documentclass[aps,reprint,showpacs,pre,superscriptaddress]{revtex4-2}
\usepackage{graphicx}

\begin{document}

\title{Brownian motion of a particle with higher-derivative dynamics}
\author{Z. C. Tu}
\email{tuzc@bnu.edu.cn}
\affiliation{Department of Physics, Beijing Normal University, Beijing 100875, China}
\affiliation{Key Laboratory of Multiscale Spin Physics, Ministry of Education, China}

\begin{abstract}
The Brownian motion of a particle with higher-derivative dynamics (HDD) coupling with a bath consisting of harmonic oscillators is investigated. The Langevin equation and corresponding Fokker-Planck equation for the Brownian motion of the HDD particle are derived. As a case study, we particularly consider a stochastic Pais-Uhlenbeck oscillator. It is found that the Boltzmann distribution is pathological while this distribution is the steady solution to the Fokker-Planck equation. [Note: This manuscript is a translation of a Chinese paper contributing for 100 anniversary of Department of Physics, Beijing Normal University.]
\end{abstract}
\maketitle

\section{introduction}

The fundamental equations of motion in classical mechanics are governed by Newton's second law (force = mass $\times$ acceleration). Acceleration is the second derivative of coordinates with respect to time, and force in classical mechanics usually does not contain derivatives higher than velocity (the first derivative of coordinates with respect to time). Thus, the fundamental equations of motion in classical mechanics are usually second order differential equations (excluding derivatives higher than second derivatives of coordinates with respect to time). Correspondingly, the Lagrangian is only a function of generalized coordinates, generalized velocities and time, which does not contain the second or higher derivatives of generalized coordinates. From the perspective of human understanding to the laws of mechanics, one can trace back to Aristotle's view of mechanics which indicates that force determines velocity, which is called the first-order dynamics in this article. We considering an object embedded in viscous fluid, the relaxation time of the momentum is given by the damping coefficient and the mass of the object. It can be proved that when the minimum resolvable time is much greater than the relaxation time, the transient process of velocity cannot be observed. In this case, Newtonian mechanics leads to a conclusion that velocity is determined by force and damping coefficients. Thus Newtonian view of mechanics degenerates into Aristotelian view of mechanics. Current research has shown that the laws of motion at the scale of microorganisms are indeed governed by Aristotelian view of mechanics. Fortunately, human beings live on the scale of meters, the resistance of air is not too large, and the inertia effect is obvious, so that human beings were able to discovery the shortcomings of Aristotelian view of mechanics and then proposed more reliable Newtonian mechanics. From the angle of theoretical development, it is natural for us to extend the fundamental equations of motion to the situation of larger spatial scales where the third or higher derivatives of coordinate with respect to time may be included. This higher-order effect is negligible at the scale of human life, but may play a role in the large scale of astronomical observations, which might provide an alternative insight to understanding certain astronomical phenomena such as dark matter deviating from current theories.

We name equations of motion containing higher (larger than second) derivatives of coordinates with respect to time as higher-derivative dynamics (HDD).
Discussions on HDD systems may be traced back to Ostrogradsky~\cite{Ostrogradsky1850}. He discussed a system where the non-degenerated Lagrangian contains second or higher derivatives of generalized coordinates with respect to time (the corresponding equations of motion contain fourth or higher derivatives). He found that the corresponding Hamiltonian has no lower bound, resulting in dynamical instability of the system. Lorentz, Abraham, and Dirac discussed the motion of electrons with radiations, and found that the force due to radiation contains third derivatives of coordinates with respect to time~\cite{Abraham1905,Dirac1938}. Dirac's theory of electron motion was generalized by Bhabha to describe the motion of neutrons~\cite{Bhabha1939}. Pais and Uhlenbeck discussed an oscillator governed by a fourth-order differential equation~\cite{Uhlenbeck1950} which became the basis for subsequent higher-derivative field theory. Chang~\cite{Chang1956} analyzed the self-acceleration behavior of Dirac's theory of electron motions and Bhabha's theory of neutron motions. He also compared the quantization equations respectively derived from the Ostrogradsky method and the Pais-Uhlenbeck method, and found that both methods lead to equivalent results~\cite{Chang1958}. As we know, the idea of HDD appears in varieties of situations such as Podolsky's generalized electrodynamics~\cite{Podolsky1942,Podolsky1944}, Polyakov's superstring theory~\cite{Polyakov86}, modified gravitational theory~\cite{Stelle1978,Simon1991,Hawking2002,Mannheim2006}, Timoshenko's beam theory~\cite{Timoshenko1921}, non-Hermitian physics~\cite{Mannheim2008,Mannheim2008b}, Starobinsky's inflation theory~\cite{Simon1992,deMedeiros2021}, and so on.

The extension of deterministic dynamics to stochastic dynamics is the other fruitful direction. At the beginning of the twentieth century, Langevin proposed a stochastic equation to describe the random motion of a Brownian particle in liquid, which became known as the Langevin equation~\cite{Langevin1908}. For the sake of simplicity, but without losing its generality, we will only mention one-dimensional motion in this paper. The Langevin equation can be expressed as
\begin{equation}\label{Langevinold}
m \frac{\mathrm{d}^2 x}{\mathrm{~d} t^2}=-\frac{\partial V}{\partial x}-\mu \frac{\mathrm{d} x}{\mathrm{d} t}+\xi(t),
\end{equation}
were $m$ and $x$ represent the mass and the position of the Brownian particle, respectively. $t$ is the time variable. $V$ represents a deterministic external field loaded on the particle. It is based on the Newton second law that Langevin intuitively wrote the above equation. The force applied on the particle by liquid molecules is phenomenologically decomposed into two terms. One is the average effect due to collisions of liquid molecules on the Brownian particle, which appears as a deterministic damping force $-\mu\mathrm{d}x/\mathrm{d}t$ with $\mu$ being a damping coefficient. The remainder fluctuating effect due to collisions of liquid molecules is expressed as a random force $\xi(t)$ which is usually assumed to be Gaussian noise with zero mean and very short-time correlation. That is,
\begin{equation}\label{Gaussian-noise1}
\langle\xi(t)\rangle =0,
\end{equation}
and
\begin{equation}\label{Gaussian-noise2}
\langle\xi(t) \xi(t')\rangle=2\mu\mathrm{k_B} T \delta(t-t'),
\end{equation}
where $\mathrm{k_B}$ is the Boltzmann factor, and $T$ is the temperature of the liquid. $\delta(t)$ represents the Dirac $\delta$-function.
The Langevin equation has became the cornerstone of stochastic thermodynamics~\cite{Sekimoto2010,Seifert2012,Peliti2021}, an emergent field of statistical physics for small systems in recent years.

How to derive the Langevin equation from the microscopic level has attracted much attention from many scientists. One of the most significant contributions is Zwanzig's scheme~\cite{Zwanzig1973,Zwanzigbook} which is essentially a generalized and simplified version of the work by Ford, Kac, and Mazur~\cite{Ford1965}. Zwanzig considered a particle with mass $m$ embedded in a bath consisting of harmonic oscillators, and the particle is linearly coupled to the harmonic oscillators. Assume that the states of harmonic oscillators initially satisfy the Boltzmann distribution at temperature $T$. When the variables of the bath were integrated, Zwanzig derived an equation of motion for the particle which exactly the Langevin equation with the noise satisfying the Gaussian form. From the perspective of theoretical extension, it is valuable to discuss the motion of a particle with HDD coupled to a bath consisting of harmonic oscillators. We will use this model as a starting point to discuss Brownian motion of HDD particles. In this work, we will derive the Langevin equation and its corresponding Fokker-Planck equation for the Brownian motion of a HDD particle. The rest of this paper is organized as follows. In Section~\ref{sec-Ostrogradsky}, we briefly introduce Ostrogradsky's construction on HDD. In Sec.~\ref{sec-Langevin-HDD}, we generalize the Langevin equation for a HDD particle using Zwanzig's scheme. In Sec.~\ref{sec-FK-HDD}, we derive the corresponding Fokker-Planck equation for a HDD particle. In Sec.~\ref{sec-Casestudy}, we discuss stochastic Pais-Uhlenbeck oscillators as a case study. A brief summary is given in Sec.~\ref{sec-sum-dis}.

\section{Ostrogradsky's construction on HDD}\label{sec-Ostrogradsky}
In this section, we introduce Ostrogradsky's approach on HDD~\cite{Ostrogradsky1850}. For convenience, we consider a system with single degree of freedom. Let $x_0$ represent the coordinate of the particle. Suppose that the Lagrangian of this system may be expressed as
\begin{equation}\label{eq-Lagrangian}
L=L(t,x_0,x_1,x_2,\cdots,x_N),
\end{equation}
where $x_{n}\equiv \mathrm{d}{x}_{n-1}/\mathrm{d}t$ ($n=1, 2,\cdots, N$) represents the $n-$th order derivative of coordinate $x_0$ with respect to time $t$.
Via the variational calculus, the Euler-Lagrange equation corresponding to Lagrangian (\ref{eq-Lagrangian}) can be derived as
\begin{equation}\label{eq-ELeq}
\sum_{n=0}^N \left(-\frac{\mathrm{d}}{\mathrm{d}t}\right)^n \frac{ \partial L}{\partial x_n } =0.
\end{equation}

Define generalized momentum
\begin{equation}\label{eq-genmomentum}
p_n=\sum_{k=n+1}^N\left(-\frac{\mathrm{d}}{\mathrm{d} t}\right)^{k-n-1} \frac{\partial L}{\partial x_k}
\end{equation}
where $n=0, 1, 2,\cdots, N-1$. In particular, when $n = 0$, the above equation leads to
\begin{equation}\label{eq-p0}
p_0=\sum_{k=1}^N\left(-\frac{\mathrm{d}}{\mathrm{d} t}\right)^{k-1} \frac{\partial L}{\partial x_k},
\end{equation}
from which we easily see that
\begin{equation}\label{eq-dp0-dt}
\frac{\mathrm{d}p_0}{\mathrm{d} t}=-\sum_{k=1}^N\left(-\frac{\mathrm{d}}{\mathrm{d} t}\right)^{k} \frac{\partial L}{\partial x_k}.
\end{equation}
Thus Euler-Lagrange equation (\ref{eq-ELeq}) can be rewritten in a new form:
\begin{equation}\label{eq-ELeq-rewrite}
\frac{\mathrm{d}p_0}{\mathrm{d} t}= \frac{\partial L}{\partial x_0}.
\end{equation}

Now let us construct the Hamiltonian. Taking $n=N-1$ in Eq.~(\ref{eq-genmomentum}), we achieve
\begin{equation}\label{eq-pN-1}
p_{N-1}= \frac{\partial L}{\partial x_N},
\end{equation}
which is a function of $t,x_0,x_1,\cdots,x_N$. From the above equation, we may solve $x_N$ in principle which is a
function of $t,x_0,x_1,\cdots,x_{N-1},p_{N-1}$ and may be formally expressed as
\begin{equation}\label{eq-xN}
x_N=\varphi\left(t, x, x_1, \ldots, x_{N-1}, p_{N-1}\right).
\end{equation}
Then, using the Legendre transformation, we obtain the Hamiltonian:
\begin{equation}\label{eq-Hamiltonian}
H=\sum_{n=0}^{N-1} p_n{x}_{n+1}-L(t,x,x_1,x_2,\cdots,x_N),
\end{equation}
where $x_N$ satisfying Eq.~(\ref{eq-xN}). Thus the Hamiltonian $H$ is a function of  $t,x_0,x_1,\cdots,x_{N-1},p_0,p_1,\cdots,p_{N-1}$.
Having the Hamiltonian, the canonical equations may be derived as
\begin{eqnarray}
&&\dot{x}_n=\frac{\partial H}{\partial p_n}=x_{n+1}\label{eq-canonical1}\\
&&\dot{p}_n=-\frac{\partial H}{\partial x_n}=\frac{\partial L}{\partial x_n}-p_{n-1},\label{eq-canonical2}
\end{eqnarray}
where $n=0, 1, 2,\cdots, N-1$ and $p_{-1}\equiv 0$. $\dot{(~)}$ represents the derivative of $(~)$ with respect to time.

\section{Langevin equation with HDD}\label{sec-Langevin-HDD}
In this section, we will disuss the Brownian motion of a particle with HDD and derive the Langevin equation with HDD following the Zwanzig scheme~\cite{Zwanzig1973,Zwanzigbook}.

Consider a large particle embedded in medium consisting of small particles. The motion of the large particle itself is governed by HDD. The motions of the small particles are governed by second derivative dynamics. For simplicity, the small particles are viewed as a bath of harmonic oscillators. The Hamiltonian of whole system is
\begin{equation}\label{eq-Hamiltonianwhole}
\mathcal{H}=H+H_B,
\end{equation}
where $H$ represents the Hamiltonian (\ref{eq-Hamiltonian}) of HDD. $H_B$ represents the Hamiltonian of the harmonic oscillators in the bath and their linear couplings with the HDD particle, which can be expressed as
\begin{equation}\label{eq-HamiltonianB}
H_B=\frac{1}{2} \sum_j\left[P_j^2+\omega_j^2\left(Q_j-\frac{\gamma_j}{\omega_j^2} x_0\right)^2\right],
\end{equation}
where $Q_j$ and $P_j$ are generalized coordinate and generalized momentum of the $j-$th oscillator, respectively. $\omega_j$ represents the frequency of the $j-$th harmonic oscillator. $\gamma_j$ represents the coupling strength of the $j-$th harmonic oscillator to the HDD particle. $x_0$ represents the coordinate of the HDD particle. Note that the mass of each oscillator has been set unit.

According to the Hamilton equations, we obtain the equations of motion for the system, which read
\begin{eqnarray}
&&\dot{Q}_j=P_j,\label{eq-dotQj}\\
&&\dot{P}_j=-\omega_j^2 Q_j+\gamma_j x_0,\label{eq-dotPj}\\
&&\dot{x}_n=x_{n+1},\label{eq-dotxn}\\
&&\dot{p}_n=\frac{\partial L}{\partial x_n}-p_{n-1}+\delta_{n 0} \sum_j \gamma_j\left(Q_j-\frac{\gamma_j}{\omega_j^2} x_0\right),\label{eq-dotpn}
\end{eqnarray}
where $n=0, 1, 2,\cdots, N-1$ and $p_{-1}\equiv 0$.

From Eqs.~(\ref{eq-dotQj}) and (\ref{eq-dotPj}), we obtain
\begin{eqnarray}
Q_j-\frac{\gamma_j}{\omega_j^2} x_0 &=&\left[Q_j(0)-\frac{\gamma_j}{\omega_j^2} x_0(0)\right] \cos \omega_j t
+P_j(0) \frac{\sin \omega_j t}{\omega_j} \nonumber\\ &-&\frac{\gamma_j}{\omega_j^2} \int_0^t \mathrm{~d} s x_1(s) \cos \omega_j(t-s),\label{eq-Qj-x0}
\end{eqnarray}
where  $x_0(0)$, $Q_j(0)$ and $P_j(0)$ are the initial values of $x_0$, $Q_j$ and $P_j$ at time $t=0$. To achieve the above result, we have used the method of integration by parts.

Substituting Eq.~(\ref{eq-Qj-x0}) into Eq.~(\ref{eq-dotpn}) with $n=0$, we arrive at
\begin{equation}\label{eq-Langeving1}
\dot{p}_0=\frac{\partial L}{\partial x_0}-\int_0^t \mathrm{d} s K(t-s) x_1(s)+\xi(t),
\end{equation}
where the kernel function $K(t)$ and the ``noise'' term $\xi(t)$ are respectively expressed as
\begin{equation}\label{eq-kernel}
K(t)  =\sum_j \frac{\gamma_j^2}{\omega_j^2} \cos \omega_j t,
\end{equation}
and
\begin{eqnarray}
\label{eq-noise}
\xi(t) &=&\sum_j \gamma_j P_j(0) \frac{\sin \omega_j t}{\omega_j}\nonumber\\
&+&\sum_j \gamma_j\left[Q_j(0)-\frac{\gamma_j}{\omega_j^2} x_0(0)\right] \cos \omega_j t.
\end{eqnarray}
Substituting expression (\ref{eq-p0}) into Eq.~(\ref{eq-Langeving1}), we arrive at a generalized Langevin equation:
\begin{equation}\label{eq-Langeving2}
\sum_{n=0}^N \left(-\frac{\mathrm{d}}{\mathrm{d}t}\right)^n \frac{ \partial L}{\partial x_n } -\int_0^t \mathrm{~d} s K(t-s) x_1(s)+\xi(t)=0.
\end{equation}

Suppose that the bath is initially in equilibrium at temperature of $T$. The states of oscillators satisfy the Boltzmann distribution $\rho_{e q}\left(Q_j(0), P_j(0)\right) \propto \mathrm{e}^{-H_B / \mathrm{k}_{\mathrm{B}} T}$. With considering this distribution and Eq.~(\ref{eq-noise}), one can arrive at Eq.~(\ref{Gaussian-noise1}) and a generalized fluctuation-dissipation relation:
\begin{equation}\label{eq-GLDR}
\langle\xi(t) \xi(s)\rangle=\mathrm{k}_{\mathrm{B}} T K(t-s).
\end{equation}
Assume that the frequency spectrum is continuous with density $g(\omega)$. Thus the summation $\sum_j$ in Eq.~(\ref{eq-kernel}) may be replaced by an integral $N_{o}\int_0^{\infty}g(\omega)$ where $N_o$ is the number of the oscillators which is usually a large number. Assume that the coupling between each oscillator and the HDD particle is constant and uniform. Thus $\gamma_j^2$ may be replaced by $\gamma N_o$. Then kernel function (\ref{eq-kernel}) may be expressed as
\begin{equation}\label{eq-kernel2}
K(t)  =\gamma\int_0^{\infty} \mathrm{d} \omega g(\omega)\frac{\cos \omega t}{\omega^2}.
\end{equation}

Taking Debye-type spectrum distribution $g=(2\mu/\gamma\pi)\omega^2$, the above kernel function is transformed into
\begin{equation}\label{eq-kerneldelta}
K(t) =2\mu\delta(t).
\end{equation}
Then Eq.~(\ref{eq-GLDR}) is degenerated into Eq.~(\ref{Gaussian-noise2}). That is, the ``noise'' term $\xi(t)$ is the Gaussian noise.
Note that $\delta(t)$ is an even function and its value is vanishing for $t\neq 0$. We can verify $ \int_0^t \mathrm{~d} s \delta(t-s) x_1(s)=\int_t^{\infty} \mathrm{d} s \delta(t-s) x_1(s)= \frac{1}{2} \int_0^{\infty} \mathrm{d} s \delta(t-s) x_1(s)=\frac{1}{2} x_1(t)$.
Thus the generalized Langevin equation (\ref{eq-Langeving2}) is degenerated into
\begin{equation}\label{eq-Langeving3}
\sum_{n=0}^N \left(-\frac{\mathrm{d}}{\mathrm{d}t}\right)^n \frac{ \partial L}{\partial x_n } -\mu x_1+\xi(t)=0.
\end{equation}
where the noise $\xi(t)$ satisfies Eqs.~(\ref{Gaussian-noise1}) and (\ref{Gaussian-noise2}). The above equation is called the Langevin equation with HDD. Relative to the Euler-Lagrange equation (\ref{eq-ELeq}), this equation has additional damping term $-\mu x_1$ and noise term $\xi(t)$. Compareing with the Langevin equation (\ref{Langevinold}), we replace $-m\mathrm{d}^2x/\mathrm{d}t^2$ and $-\partial V/\partial x$ with $\sum_{n=1}^N (-{\mathrm{d}}/{\mathrm{d}t})^n { \partial L}/{\partial x_n } $ and ${ \partial L}/{\partial x_0 }$ , respectively.
It is not hard to verify that, for the case of $N=1$ with $L=(m/2)x_1^2-V(t,x_0)$, where the dynamics is second-order (i.e., Newtonian mechanics), Eq.~(\ref{eq-Langeving3}) is degenerated into Eq.~(\ref{Langevinold}).

\section{Fokker-Planck equation with HDD}\label{sec-FK-HDD}
As we know, there is a Fokker-Planck equation corresponding to the Langevin equation (\ref{Langevinold})  in non-equilibrium statistical mechanics, which describes the evolution of the distribution function in phase space. In this section, we will derive the Fokker-Planck equation corresponding to the Langevin equation (\ref{eq-Langeving3})  with HDD.

Similar to the discussion in the above section, Eq.~(\ref{eq-Langeving1}) can be transformed into
\begin{equation}\label{eq-Langevin3}
\dot{p}_0=\frac{\partial L}{\partial x_0}-\mu x_1+\xi(t),
\end{equation}
which reminds us that Eq.~(\ref{eq-dotpn}) can be rewritten as
\begin{equation}\label{eq-pndot3}
\dot{p}_n=\frac{\partial L}{\partial x_n}-p_{n-1}+\delta_{n 0}\xi(t),
\end{equation}
where we have redefined $p_{-1}\equiv \mu x_1$.
The above equation and Eq.~(\ref{eq-dotxn}) describe the evolution of microscopic trajectories in phase space $\{\mathbf{\Gamma}\}=\{(x_0,x_1,\ldots,x_{N-1},p_0,p_1,\ldots,p_{N-1})\}$.

Let $f(t,\mathbf{\Gamma})\mathrm{d}\mathbf{\Gamma}$ represent the probability of the HDD particle appearing within the region between $\mathbf{\Gamma}$ and $\mathbf{\Gamma}+\mathrm{d}\mathbf{\Gamma}$ in the phase space. According to the conservation law of probability flow~\cite{Reichl1998}, we have
\begin{equation}\label{eq-conservation1}
\frac{\partial f}{\partial t}=-\sum_{n=0}^{N-1}\left[\frac{\partial({\dot{x}}_nf)}{\partial x_n}+\frac{\partial({\dot{p}}_nf)}{\partial p_n}\right].
\end{equation}
Substituting Eqs.~(\ref{eq-dotxn}) and (\ref{eq-pndot3}) into the above equation, we arrive at
\begin{eqnarray}
\frac{\partial f}{\partial t}=&-&\sum_{n=0}^{N-1}\left\{\frac{\partial}{\partial x_n}(x_{n+1}f)+\frac{\partial}{\partial p_n}\left[\left(\frac{\partial L}{\partial x_n}-p_{n-1}\right)f\right]\right\} \nonumber\\
&-&\xi(t)\frac{\partial f}{\partial p_0}\label{eq-conservation2}
\end{eqnarray}

The observable distribution function of the HDD particle in the phase space is defined as the average of $f$ with respect to the noise, that is,
\begin{equation}\label{eq-distribution}\rho(t,\mathbf{\Gamma})=\langle f(t,\mathbf{\Gamma})\rangle_\xi.\end{equation}
Following Reichl's derivation~\cite{Reichl1998} of the Fokker-Planck equation from the Langevin equation, we can derive the Fokker-Planck equation corresponding to the Langevin equation (\ref{eq-Langeving3}) [equivalently, Eqs.~(\ref{eq-dotxn}) and (\ref{eq-pndot3})] with HDD from Eqs.~(\ref{eq-conservation2}) and (\ref{eq-distribution}). The derived Fokker-Planck equation with HDD is
\begin{eqnarray}
\frac{\partial \rho}{\partial t}=&-&\sum_{n=0}^{N-1}\left\{\frac{\partial}{\partial x_n}(x_{n+1}\rho)+\frac{\partial}{\partial p_n}\left[\left(\frac{\partial L}{\partial x_n}-p_{n-1}\right)\rho\right]\right\} \nonumber\\
&+&\mu k_BT \frac{\partial^2 \rho}{\partial p_0^2}\label{eq-FK2}
\end{eqnarray}

As an example, we consider the case of $N=1$ with $L=(m/2)x_1^2-V(t,x_0)$. The generalized momentum is $p_0={\partial L}/{\partial x_1}=mx_1$. Then $x_1=p_0/m$ and $p_{-1}=\mu x_1=\mu{p_0}/{m}$. Thus the above Fokker-Planck equation~(\ref{eq-FK2}) with HDD is degenerated into
the well-known Fokker-Planck equation,
\begin{equation}\label{eq-FK-old}
\frac{\partial \rho}{\partial t}=\frac{p_0}{m}\frac{\partial \rho}{\partial x_0}-\frac{\partial}{\partial p_0}\left[\left(\frac{\partial V}{\partial x_0}+\mu\frac{p_0}{m}\right) \rho\right] +\mu k_BT \frac{\partial^2 \rho}{\partial p_0^2},
\end{equation}
corresponding to the Langevin equation~(\ref{Langevinold}).

\section{Stochastic Pais-Uhlenbeck oscillator}\label{sec-Casestudy}
In this section, we will discuss stochastic Pais-Uhlenbeck oscillators as a case study.
The Lagrangian of the Pais-Uhlenbeck oscillator can be expressed as~\cite{Uhlenbeck1950,Mannheim2008}
\begin{equation}\label{eq-LPU}
L=\frac{Y}{2}[x_2^2-(\omega_1^2+\omega_2^2) x_1^2+\omega_1^2 \omega_2^2 x_0^2],
\end{equation}
where frequencies $\omega_1$ and $\omega_2$ are independent of time. $Y$ is a constant. The above Lagrangian corresponds to the case of $N=2$, thus the corresponding equaiton of motion is fourth-order. Substituting equation (\ref{eq-LPU}) into Eq.~(\ref{eq-Langeving3}), we obtain a fourth-order Langevin equation
\begin{equation}\label{eq-Lang-PU}
Y\left[\frac{\mathrm{d}^4 x_0}{\mathrm{d} t^4}+(\omega_1^2+\omega_2^2) \frac{\mathrm{d}^2 x_0}{\mathrm{d} t^2}+\omega_1^2 \omega_2^2 x_0\right]-\mu \frac{\mathrm{d} x_0}{\mathrm{d} t}+\xi(t)=0
\end{equation}
which contains both the damping term $-\mu {\mathrm{d} x_0}/{\mathrm{d} t}$ and the noise term $\xi(t)$. A oscillator described in the above equation is named as stochastic Pais-Uhlenbeck oscillator. To the best of our knowledge, Nesterenko first discussed the Pais-Uhlenbeck oscillator with a damping term~\cite{Nesterenko2007}, while Urenda-C\'{a}zares et al. discussed the Pais-Uhlenbeck oscillator with a noise term~\cite{Urenda2019}. They wrote the corresponding equations phenomenologically without microscopic derivations. In the above discussion, we have made up for this shortage.

According to Eq.~(\ref{eq-pN-1}), we obtained $p_1=Yx_2$. So  $x_2=p_ 1/Y$ and $L=p_1^2/(2Y)+(Y/2)[\omega_1^2 \omega_2^2 x_0^2-(\omega_1^2+\omega_2^2) x_1^2]$. Substituting it into Eq.~(\ref{eq-FK2}), we derive the Fokker-Planck equation
\begin{eqnarray}
\frac{\partial \rho}{\partial t}&=&-x_1 \frac{\partial \rho}{\partial x_0}-\frac{p_1}{Y} \frac{\partial \rho}{\partial x_1}-(Y \omega_1^2 \omega_2^2 x_0-\mu x_1) \frac{\partial \rho}{\partial p_0}\nonumber\\ &+&[Y(\omega_1^2+\omega_2^2) x_1+p_0] \frac{\partial \rho}{\partial p_1}+\mu \mathrm{k}_{\mathrm{B}} T \frac{\partial^2 \rho}{\partial p_0^2}\label{eq-FP-PUoscillator}
\end{eqnarray}
for the stochastic Pais-Uhlenbeck oscillator.

According to Eq.~(\ref{eq-Hamiltonian}), the Hamiltonian of the Pais-Uhlenbeck oscillator can be written as
\begin{equation}\label{eq-PU-Hamiltonian}
H=p_0 x_1+\frac{p_1^2}{2 Y}+\frac{Y}{2}[(\omega_1^2+\omega_2^2) x_1^2-\omega_1^2 \omega_2^2 x_0^2].
\end{equation}
It is not hard to verify that the Boltzmann distribution $\rho_{s}\propto \mathrm{e}^{-H/\mathrm{k_B}T}$ is the solution to the Fokker-Planck equation (\ref{eq-FP-PUoscillator}). Thus, although the Hamiltonian (\ref{eq-PU-Hamiltonian}) has no lower bound, there is still a steady-state distribution $\rho_{s}\propto \mathrm{e}^{-H/\mathrm{k_B}T}$ for the stochastic Pais-Uhlenbeck oscillator. This distribution is pathological if no specific constraint is applied on the Hamiltonian. In addition, whether this distribution is stable still needs to be further explored.

\section{Summary and discussion}\label{sec-sum-dis}
In the above discussion, we have investigated the motion of a HDD particle coupled with a bath consisting of harmonic oscillators, and then derived the Langevin equation (\ref{eq-Langeving3}) and its corresponding Fokker-Planck equation (\ref{eq-FK2}) for the HDD particle. As a case study, we consider the stochastic Pais-Uhlenbeck oscillator and present the corresponding Langevin equation (\ref{eq-Lang-PU}) and Fokker-Planck equation (\ref{eq-FP-PUoscillator}). These equations can be used as a starting point for studying the Brownian motion of HDD particles. It should be noted that in the above discussion, the dynamics of harmonic oscillators in the bath is still second-order. If the dynamics of harmonic oscillators is replaced with the HDD, such as the Pais-Uhlenbeck oscillators, we need further discuss whether the above results can still be maintained.
In addition, we have only considered the one-dimensional case where all Lagrangians are non-degenerated. As a result, there is no one-dimensional odd-order-derivative dynamics in our theoretical framework. For 2-dimensional or higher-dimensional cases where Lagrangians might be degenerated~\cite{Motohashi2015}. We expect that the forms of the Langevin equation and the Fokker-Planck equation are not essentially changed for the degenerated cases.

Stochastic thermodynamics~\cite{Sekimoto2010,Seifert2012,Peliti2021} is a frontier research field emerged in recent years for describing thermodynamic behaviors of small systems, which is mainly based on the conventional Langevin equation (\ref{Langevinold}) and its corresponding Fokker-Planck equation (\ref{eq-FK-old}). With the aid of these two equations, the quantities of work, heat, and entropy can be well defined on microscopic trajectories. In this paper, we have derived the Langevin equation (\ref{eq-Langeving3}) and its corresponding Fokker-Planck equation (\ref{eq-FK2}) for a HDD Brownian particle. It is valuable for us to extend current stochastic thermodynamics to the case of HDD based on these two equations. The crucial matter lies in the proper definitions of work, heat, and entropy. We expect that their definitions are similar to those in current stochastic thermodynamics~\cite{Sekimoto2010,Seifert2012,Peliti2021}.

Different from the Langevin equation and Fokker-Planck equation, the path integral method offers another way to describe stochastic processes. We have noted that Kleinert~\cite{Kleinert86} discussed the path integral for the Pais-Uhlenbeck oscillator, and that Dean et al.~\cite{Miaobing2019} discussed the path integral of HDD in quadratic form. In both cases, they achieved analytical solutions of propagators. Although it is difficult to obtain analytical solutions in general cases, we may still borrow the idea of path integrals to extend the framework of stochastic thermodynamics to the situation of HDD. We will further discuss this point in the follow-up work.

Note: The Chinese version of this manuscript is referred to Ref.~\cite{TuZCPhys100}.

\section*{ACKNOWLEDGEMENTS}
The author would like to thank Xiuhua Zhao and Yating Wang for their careful reading the manuscript. This work is supported by the National Natural Science Foundation of China (Grant No. 11975050).

\end{document}